\documentclass[pra,preprint]{revtex4-1}

\usepackage{amsmath}  
\usepackage{amsfonts} 
\usepackage{graphicx}

\begin{document}

\title{Optimal number of terms in QED series and its consequence in condensed matter implementations of QED}

\author{Eugene B. Kolomeisky}
\email{ek6n@virginia.edu}

\affiliation
{Department of Physics, University of Virginia, P. O. Box 400714,
Charlottesville, Virginia 22904-4714, USA}
\date{\today}

\begin{abstract}
In 1952 Dyson put forward a simple and powerful argument indicating that the perturbative expansions of QED are asymptotic.  His argument can be related to Chandrasekhar's limit on the mass of a star for stability against gravitational collapse.  Combining these two arguments we estimate the optimal number of terms of the QED series to be $3.1(137)^{3/2}\approx5000$.  For condensed matter manifestations of QED in narrow band-gap semiconductors and Weyl semimetals the optimal number of terms is around $80$ while in graphene the utility of the perturbation theory is severely limited.   
\end{abstract}


\maketitle

\section{Introduction}

Quantum electrodynamics (QED) calculates measurable quantities through a power series that expands in powers of the fine structure constant $\alpha=e^{2}/\hbar c$.  Over sixty years ago Dyson \cite{Dyson} gave an argument indicating that QED series are asymptotic (i.e. there exists an optimal number of terms that best approximates the physical quantity being computed; keeping more terms does not improve but rather worsens the accuracy).  Dyson argued that this optimal number  is of the order $1/\alpha\approx137$.   Migdal and Krainov \cite{MK} reconsidered Dyson's argument and arrived at a different optimal number of terms of the order $1/\alpha^{3/2}\approx 137^{3/2}$.  Both estimates predict a very large number thus explaining why in practice the asymptotic character of QED series does not limit the utility of the perturbation theory. The goal of this paper is two-fold:

(i)  First, we demonstrate how Dyson's argument can be made quantitative. This allows us to identify the origin of discrepancy between the two predictions \cite{Dyson,MK} and improve the accuracy of the Migdal-Krainov estimate.  The latter is made possible by reducing the argument to mathematically identical problem of stability of a star against gravitational collapse \cite{LL5}.

(ii)  Second, we point out that QED is not the only context to which Dyson's argument can be applied.  Certain condensed matter systems, specifically narrow band-gap semiconductors \cite{Keldysh}, Weyl semimetals \cite{AB} and graphene \cite{graphene} in their low-energy limit, mimic QED with vastly different effective fine structure constants \cite{KSZ}.  We show that in these systems the asymptotic character of appropriate perturbative series is not necessarily of purely theoretical interest.  Specifically, in graphene the utility of the perturbation theory is severely limited.   

\section{Dyson's argument in QED and gravitational collapse connection}

In order to demonstrate that physical quantities regarded as functions of $\alpha$ or equivalently $e^{2}$ have a singularity at $e^{2}=0$ Dyson argued by contradiction.  If we assume that there is no singularity for $e^{2}=0$, the value of any physical quantity at $e^{2}=0$ should not depend on the way $e^{2}$ approaches zero -- whether it is from the $e^{2}>0$ or $e^{2}<0$ domains.  For $e^{2}>0$ the QED vacuum as a particle-free ground state is clearly stable.   On the other hand, if $e^{2}<0$ the particle-free vacuum is unstable and thus does not represent the lowest energy state of the system.  To clarify the meaning of the vacuum instability, Dyson interpreted the $e^{2}<0$ situation as describing a fictitious world where like charges attract and unlike charges repel.  In such a world were a quantum fluctuation to create $N$ electron-positron pairs, the repulsive electrons and positrons would spatially separate while the attractive particles of the same type would segregate together.  For $N\gg1$ the total energy of the well separated electron and positron regions will become lower than the energy of the particle-free vacuum because the energy penalty in creating pairs (proportional to their number $N$), will be inevitably compensated by the energy gain of attraction between like particles (proportional to the number of pair interactions $N(N-1)\approx N^{2}$).  Thus there is a smallest number of pairs $N_c$ beyond which the vacuum is unstable with respect to creation of another pair.  Since the vacuum is stable in the real $e^{2}>0$ world and unstable in the fictitious $e^{2}<0$ world, $e^{2}=0$ must be a singular point.  

Dyson commented that the divergence of the QED series in the real world is associated with virtual processes in which large numbers of particles are involved.  In the QED perturbation series terms of the order $\alpha^{N}\propto e^{2N}$ arise from processes involving $N$ virtual electron-positron loops which, as was argued by Migdal and Krainov \cite{MK}, correspond to $N$ electron-positron pairs in the fictitious $e^{2}<0$ world.  In other words, creation of $N$ electron-positron pairs in the $e^{2}<0$ world corresponds to the $N$th order of the perturbation theory in the physical world, suggesting that the critical number of pairs $N_{c}$ becomes the optimal number of terms of asymptotic QED series.

Even though the instability of the $e^{2}<0$ vacuum has its origin in Coulomb attraction of like charges overcoming the rest energy of pairs ($2Nmc^{2}$), the critical pair number $N_{c}$ does not depend on the electron mass $m$.  This a consequence of the quantum nature of the problem (which puts a lower bound on the energy of a bound pair), and can be readily inferred from dimensional analysis:

The problem is fully specified by the dimensionless parameters $N$, $\alpha$, and by the electron Compton wavelength $\lambda=\hbar/mc$.  If there exists a critical number of pairs $N_{c}$, it can only be a function of the remaining independent dimensionless parameters of the problem.  However there is only one length scale $\lambda$ available, so that no dimensionless quantity can depend on it.  Therefore $N_{c}$ cannot depend on $\lambda$, and thus is independent of the electron mass $m$;  the only possible outcome is $N_{c}=f(\alpha)$, where $f$ is a yet unknown function which diverges in the non-interacting $\alpha\rightarrow 0$ limit.  

Dyson argued that a state whose energy is lower than that of the particle-free $e^{2}<0$ vacuum can be constructed from well-separated regions of electrons and positrons "without using particularly small regions or high charge densities, so that the validity of the classical Coulomb potential is not in doubt" \cite{Dyson}.  However, relativistic treatment of the particle dynamics is necessary, so that the Hamiltonian describing one of the regions occupied by $N$ like particles will be chosen in the semi-relativistic form 
\begin{equation}
\label{Hamiltonian}
H=c\sum_{i=1}^{N}\sqrt{p_{i}^{2}+(mc)^{2}}-e^{2}\sum_{i<j}\frac{1}{|\textbf{r}_{i}-\textbf{r}_{j}|}
\end{equation}
i. e. the particles obey the relativistic dispersion relation but attract each other according to the classical non-relativistic Coulomb law.  The Hamiltonian (\ref{Hamiltonian}) is implicit in Refs. \cite{Dyson,MK}. Its quantum-mechanical  ground state is a result of the interplay between the degeneracy pressure of confined fermions that tends to expand the region and the $1/r$ attraction among them that causes contraction.  A mathematically identical situation is encountered in the study of the gravitational equilibrium of bodies of large mass \cite{LL5} where it is known that there is the Chandrasekhar-Landau (CL) mass limit, beyond which the degeneracy pressure can no longer prevent gravitational collapse. The CL mass limit directly translates into an expression for the optimal number of terms $N_{c}$ of the QED series as explained below.  The relevance of the semi-relativistic Hamiltonian of the form (\ref{Hamiltonian}) to the problem of gravitational collapse was realized by L\'evy-Leblond \cite{Leblond} whose analysis we now follow.      

If $p$ is an average momentum of the particle while $r$ is an average distance between the particles, in the $N\gg1$ limit the total energy of the interacting system described by the Hamiltonian (\ref{Hamiltonian}) can be estimated as 
\begin{equation}
\label{general_energy estimate}
E(N)\simeq Nc\sqrt{p^{2}+(mc)^{2}}-\frac{N^{2}e^{2}}{r}.
\end{equation}
The Pauli principle requires that there not be more than one fermion per de Broglie wavelength $\hbar/p$.  Then $N$ particles will occupy the total volume of about $N(\hbar/p)^{3}$ and the average distance between the particles (and the size of the region) is of the order $r\simeq N^{1/3}\hbar/p$ which transforms the expression (\ref{general_energy estimate}) into 
\begin{equation}
\label{E_of_p}
E(N)\simeq Nmc^{2}\left (\sqrt{x^{2}+1}-\alpha N^{2/3} x\right ),~~~x=\frac{p}{mc}.
\end{equation}
Since the energy per particle depends on the particle number $N$ through the $\alpha N^{2/3}$ combination and the critical number $N_{c}$ can only depend on $\alpha$, this gives the estimate of Migdal and Krainov, $N_{c}\simeq \alpha^{-3/2}$, whose origin is in the fermion nature of the particles.  The original analysis of Migdal and Krainov is similar except that they treat the cases of non-relativistic $x\ll1$ and ultra-relativistic $x\gg1$ particles separately.   

On the other hand, if the particles were bosons, the average distance between them would be dictated by the uncertainty principle, $r\simeq\hbar/p$.  Substituting this into Eq.(\ref{general_energy estimate}) would give an expression similar to (\ref{E_of_p}) except that $\alpha N^{2/3}$ would be replaced by $\alpha N$.  This reproduces Dyson's estimate $N_{c}\simeq \alpha^{-1}$ whose origin is rooted in the bosonic nature of the particles.  Rigorous analysis of the Hamiltonian (\ref{Hamiltonian}) conducted by Lieb and Yau \cite{Lieb} confirms the relevance of the $\alpha N^{2/3}$ or $\alpha N$ combinations for fermions or bosons, respectively.

As a function of the free parameter $x$ the function (\ref{E_of_p}) has a minimum for $N<N_{c}\simeq \alpha^{-3/2}$ while for $N>N_{c}$ the minimum does not exist and the energy can be made as negative as needed by choosing large enough $x$:  the Hamiltonian (\ref{Hamiltonian}) is unbounded from below and the system faces inevitable collapse analogous to the gravitational collapse \cite{LL5}.  Indeed, whenever the minimum of (\ref{E_of_p}) exists, the lowest energy of the two well-separated electron-positron regions and the size of each region $r_{0}$ can be estimated as
\begin{equation}
\label{lowest_energy}
2E_{0}\simeq2Nmc^{2}\left [1-\left (\frac{N}{N_{c}}\right )^{4/3}\right ]^{1/2}
\end{equation} 
and 
\begin{equation}
\label{size}
r_{0}\simeq \lambda N_{c}^{2/3}N^{-1/3}\left [1-\left (\frac{N}{N_{c}}\right )^{4/3}\right ]^{1/2}
\end{equation}
respectively.  Since the energy (\ref{lowest_energy}) is counted with respect to that of the particle-free vacuum, the states with $N<N_{c}$ have higher energy than that of the original vacuum and the latter is stable.  Therefore creation of the electron-positron pairs is not energetically favorable until $N$ reaches $N_{c}$.  This is when both $2E_{0}$ and $r_{0}$ vanish with the latter being analogous to gravitational collapse \cite{LL5}.  Specifically the dependence of the size of the region on the number of particles it contains (\ref{size}) parallels the dependence of the radius of gravitating region on its mass \cite{LL5}.   The correspondence between the $N=N_{c}$ condition with that of gravitational collapse can be further coraborated by looking at the ultra-relativistic (or $m=0$) limit when the estimate (\ref{E_of_p}) becomes
\begin{equation}
\label{ultra}
E'(N) \simeq Ncp\left [1-\left (\frac{N}{N_{c}}\right )^{2/3}\right ].
\end{equation} 
We now see that $N=N_{c}$ corresponds to the state of neutral equilibrium when an arbitrary region size is allowed.  Incidentally, neutrality of the equilibrium guaranties that $N_{c}$ can be determined from the semi-relativistic Hamiltonian (\ref{Hamiltonian}) with Coulomb interaction chosen in the classical non-relativistic form.  For $N>N_{c}$ the region would catastrophically contract with the energy decreasing without bound which corresponds to rapid disintegration of the vacuum.  Since the ultra-relativistic case is exactly solvable and $N_{c}$ is mass-independent, the value of $N_{c}$ follows from the expression for the CL mass limit \cite{LL5}
\begin{equation}
\label{CL_limit}
N_{c}=3.1\alpha^{-3/2}\approx 5000.
\end{equation}
This improves on the Migdal-Krainov estimate of the optimal number of terms in QED series and demonstrates once again that asymptotic nature of QED series is only of purely theoretical interest.  The first proof that the CL gravitational collapse formulas are correct up to a reasonable numerical constant was given by Lieb and Thirring \cite{LT}.

\section{Dyson's argument for narrow band-gap semiconductors and Weyl semimetals}

QED is not the only context to which the Dyson argument is relevant.  Indeed, it is known that the low-energy physics of narrow band-gap semiconductors \cite{Keldysh} mimics that of QED \cite{ZP}.  Here excitation of an electron-hole pair parallels the creation of an electron-positron pair in QED, with the band gap representing the combined rest energy of the particles.  In a two-band approximation (well obeyed in narrow band-gap semiconductors of the $InSb$ type) the low-energy electron (hole) dispersion law is pseudo-relativistic \cite{Keldysh}
\begin{equation}
\label{dispersion}
\varepsilon(\textbf{p})=\pm v_{F}\sqrt{p^{2}+(m_{b}v_{F})^{2}}.
\end{equation}            
Here the upper and lower signs correspond to the conduction and valence bands, respectively, $2m_{b}v_{F}^{2}$ is the energy band gap that parallels twice the rest energy of a particle of mass $m_{b}$, the effective band electron mass, and $v_{F}$ is the velocity of a high-momentum particle analogous to the speed of light $c$.  In the special case of zero band gap, $m_{b}=0$, Eq.(\ref{dispersion}) also describes Weyl semimetals \cite{AB}.  Application of Dyson's argument to these systems entails studying the ground-state properties of the Hamiltonian
\begin{equation}
\label{nbgsHamiltonian}
H=v_{F}\sum_{i=1}^{N}\sqrt{p_{i}^{2}+(m_{b}v_{F})^{2}}-\frac{e^{2}}{\epsilon}\sum_{i<j}\frac{1}{|\textbf{r}_{i}-\textbf{r}_{j}|}
\end{equation}
where $\epsilon$ is material's dielectric constant.  Apart from notational differences  $v_{F}\rightarrow c$, $m_{b}\rightarrow m$ and $e^{2}/\epsilon \rightarrow e^{2}$, the Hamiltonians (\ref{Hamiltonian}) and (\ref{nbgsHamiltonian}) are mathematically identical, and the results (\ref{lowest_energy})-(\ref{CL_limit}) apply to the narrow band-gap semiconductors and Weyl semimetals with the provision that QED's fine structure constant $\alpha$ is to be replaced by material's fine structure \begin{equation}
\label{cm_fine_structure_constant}
\beta=\frac{e^{2}}{\hbar v_{F}\epsilon}.
\end{equation}   
Compared to their vacuum electron-positron counterparts, electrons and holes in narrow band-gap semiconductors are very light ($m_{b}\simeq 0.01m$) and the limiting velocity $v_{F}$ is nearly three orders of magnitude smaller than the speed of light ($v_{F}\approx 4.3 \times 10^{-3}c$) \cite{Zawadzki}.  Choosing $\epsilon=15$ which is typical for narrow band-gap semiconductors \cite{Kittel} we find $\beta=0.11$ which is more than an order of magnitude larger than QED's fine structure constant.  Correspondingly, the optimal number of terms of the perturbation theory in $\beta$ in narrow band-gap semiconductors is $N_{c}\approx 80$ which is substantially smaller than its QED counterpart (\ref{CL_limit}).  

Weyl semimetals are expected to have material parameters, similar to those of narrow band-gap semiconductors: $v_{F}\simeq 10^{-2}c$ and $\epsilon \simeq 10$ \cite{AB}.  This means that their fine structure constant is about the same, $\beta \simeq 0.1$, thus implying that $N_{c} \simeq 80$.

\section{Dyson's argument for graphene}

Although the asymptotic character of perturbative series in narrow band-gap semiconductors and Weyl semimetals is substantially more pronounced than in QED, this is still of no practical relevance as $N_{c} \simeq 80 \gg 1$.  We now give an example of experimentally available condensed matter system - graphene - where Dyson's argument sets practical limitations on the utility of the perturbation theory.  First application of Dyson's argument to graphene was given in Ref. \cite{Das1};  for the special case of suspended graphene our conclusions below differ from those of Ref. \cite{Das1}.  

Graphene is a two-dimensional semimetal with linear in momentum dispersion relationship (see Eq.(\ref{dispersion}) with $m_{b}=0$) embedded into a three-dimensional space thus implying that electrons and holes confined to two dimensions interact with each other according to the three-dimensional Coulomb law.  Application of Dyson's argument to graphene entails studying ground-state properties of the Hamiltonian  
\begin{equation}
\label{graphene_Hamiltonian}
H=v_{F}\sum_{i=1}^{N}|\textbf{p}_{i}|-\frac{e^{2}}{\epsilon}\sum_{i<j}\frac{1}{|\textbf{r}_{i}-\textbf{r}_{j}|}.
\end{equation}
Since the particles are now confined to two dimensions, the results (\ref{lowest_energy})-(\ref{CL_limit}) are not applicable to graphene.  Their counterparts can be found by proceeding along the same lines as before. Indeed, the ground-state energy of $N\gg 1$ massless fermions interacting according to the Hamiltonian (\ref{graphene_Hamiltonian}) can be estimated as 
\begin{equation}
\label{general_graphehe_energy_estimate}
E'(N)\simeq Nv_{F} p - \frac{N^{2}e^{2}}{\epsilon r}.
\end{equation}
The Pauli principle requires that $N$ particles will occupy the total area of about $N(\hbar/p)^{2}$, and the average distance between the particles (and the size of the region) is of the order $r\simeq N^{1/2}\hbar/p$ which transforms the expression (\ref{general_graphehe_energy_estimate}) into 
\begin{equation}
\label{graphene_lowest_energy}
E'(N)\simeq N v_{F}p\left [1-\left (\frac{N}{N_{c}}\right )^{1/2}\right ]
\end{equation}
resembling Eq.(\ref{ultra}) with 
\begin{equation}
\label{graphene_limit}
N_{c}=0.7 \beta^{-2},
\end{equation}
the graphene counterpart of Eq.(\ref{CL_limit}), having the meaning of the optimal number of terms of the perturbation theory.  The numerical factor of $0.7$ was borrowed from Ref. \cite{Das1} where explicit solution of the two-dimensional collapse problem was given.   

With $e^{2}/\hbar v_{F}\approx2.5$ \cite{graphene} the graphene fine structure constant is $\beta \approx 2.5/\epsilon$.  If the screening is solely due to graphene's electrons (suspended graphene) one should use the value of $\epsilon=5$ \cite{RPA}.  Then $\beta\approx 0.5$ and the optimal number of terms estimated from (\ref{graphene_limit}) is $N_{c}\approx 3$.  At this point we recall that the theory developed here is applicable provided $N\gg1$.  Having found $N_{c}\approx 3$ means that one is close to the verge of applicability of the theory and that the actual number should not be taken literally.  The consequence is that in suspended graphene the utility of the perturbation theory in $\beta$ is at best limited to only few lowest order terms.  This is in variance with conclusion of Ref.\cite{Das1} which employed $\epsilon =1$; then $\beta \approx 2.5$ thus implying that the perturbation theory in the fine structure constant is never applicable.    

For the case of graphene on quartz one should choose $\epsilon\approx2.5$ \cite{Ando}.  Then one finds $\beta\approx 1$ and $N_{c}\approx 0.7$ with the implication that the perturbation theory in $\beta$ is useless.  

An alternative and reliable expansion in (large) number of Dirac fermion species capable of handling strong coupling problems like those offered by graphene was recently put forward in Ref.\cite{Das2}. 

\begin{acknowledgments}

The author is grateful to P. Arnold and J. P. Straley for comments and to S. Das Sarma for bringing Refs. \cite{Das1} and \cite{Das2} to author's attention.  This work was supported by US AFOSR Grant No. FA9550-11-1-0297.

\end{acknowledgments}


\begin{thebibliography}{17}

\bibitem{Dyson}  F. J. Dyson, Phys. Rev. \textbf{85}, 631 (1952).

\bibitem{MK}  A. B. Migdal and V. P. Krainov, \textit{Approximate Methods of Quantum Mechanics} (NEO Press, Ann Arbor, 1968), Chapter 1.4;  A. B. Migdal, \textit{Qualitative Methods in Quantum Theory} (Westview Press, 2000), Chapter 1.3.

\bibitem{LL5}  L. D. Landau and E. M. Lifshitz, Statistical Physics, vol.V, Part 1, (Pergamon, 1980), Section 107, specifically Eq.(107.17).

\bibitem{Keldysh}  L. V. Keldysh, Zh. Eksp. Teor. Fiz. \textbf{45}, 364 (1963) [Sov. Phys. JETP \textbf{18}, 253 (1964)];  see also P. A. Wolff, J. Phys. Chem. Sol. \textbf{25}, 1057 (1964);  A. G. Aronov and G. E. Pikus, Zh. Eksp. Teor. Fiz. \textbf{51}, 281 (1966) [Sov. Phys. JETP \textbf{24}, 188 (1967)], and  M. H. Weiler, W. Zawadzki, and B. Lax, Phys. Rev. \textbf{163}, 733 (1967).   

\bibitem{AB}  A. A. Abrikosov and S. D. Beneslavski\u{i}, Zh. Eksp. Teor. Fiz. \textbf{59}, 1280 (1970) [Sov. Phys. JETP \textbf{32}, 699 (1971)]; J. Low Temp. Phys. \textbf{5}, 141 (1971);  X. Wan, A.M. Turner, A. Vishwanath and S.Y. Savrasov, Phys. Rev. B \textbf{83}, 205101 (2011); V. Aji, arXiv:1108.4426;  A.A. Burkov and L. Balents,  Phys. Rev. Lett. \textbf{107}, 127205 (2011); A.A. Burkov, M.D. Hook, L. Balents, Phys. Rev. B \textbf{84}, 235126 (2011).

\bibitem{graphene}  A. H. Castro Neto, F. Guinea, N. M. R. Peres, K. S. Novoselov and A. K. Geim, Rev. Mod. Phys. \textbf{81}, 110 (2009), and references therein.

\bibitem{KSZ}  E. B. Kolomeisky, J. P. Straley and H. Zaidi, Phys. Rev. B \textbf{88}, 165428 (2013), and references therein.

\bibitem{Leblond}  J.-M. L\'evy-Leblond, J. Math. Phys. \textbf{10}, 806 (1969).

\bibitem{Lieb}  E. H. Lieb and H.-T. Yau,  Commun. Math. Phys. \textbf{112}, 147 (1987).

\bibitem{LT}  E.H. Lieb and W. Thirring,  Annals of Phys. (N.Y.) \textbf{155}, 494 (1984).

\bibitem{ZP}  Ya. B. Zel'dovich and V. S. Popov, Usp. Fiz. Nauk \textbf{105}, 403 (1971) [Sov. Phys. Uspekhi \textbf{14}, 673 (1972)], and references therein;  V. S. Popov, Yad. Fiz. \textbf{64}, 421 (2001) [Phys. At. Nucl., \textbf{64}, 367 (2001)], and references therein.   

\bibitem{Zawadzki}  W. Zawadzki, Phys. Rev. B \textbf{72}, 085217 (2005);  W. Zawadzki, in \textit{Optical Properties of Solids}, edited by E. D. Heidemenakis (Gordon and Breach, New York, 1970), p. 179.

\bibitem{Kittel}  C. Kittel, \textit{Introduction to Solid State Physics}, 7th Edition (John Wiley \& Sons, Inc., New York, 1996), Chapter 8.

\bibitem{Das1}  E. Barnes, E. H. Hwang, R. E. Throckmorton, and S. Das Sarma, Phys. Rev. B \textbf{89}, 235431 (2014).

\bibitem{RPA}  J. Gonz\'alez, F. Guinea, and M. A. H. Vozmediano, Nucl. Phys. B \textbf{424}, 595 (1994); J. Low Temp. Phys. \textbf{99}, 287 (1995).

\bibitem{Ando}  T. Ando, J. Phys. Soc. Japan \textbf{75},  074716 (2006).

\bibitem{Das2}  J. Hofmann, E. Barnes, and S. Das Sarma, Phys. Rev. Lett. \textbf{113}, 105502 (2014).

\end{thebibliography}
\end{document}